\documentstyle[11pt,newpasp,twoside,epsf]{article}
\def\ee #1 {\times 10^{#1}}          
\def\ut #1 #2 { \, \textrm{#1}^{#2}} 
\def\u #1 { \, \textrm{#1}}          

\def\deg    {$^{\circ}$}                        
\def\kms    {\hbox{km{\hskip0.1em}s$^{-1}$}}    
\def\msol   {\hbox{$M_\odot$}}                  

\def\etal   {{\it et al. }}                     

\def\edcomment#1{\iffalse\marginpar{\raggedright\sl#1\/}\else\relax\fi}
\reversemarginpar

\begin{document}
\title{Supernova OH (1720 MHz) Masers in Sgr A East, W28 and G359.1-0.5}
\author{F. Yusef-Zadeh}
\affil{Department of Physics and Astronomy, Northwestern University,
Evanston, IL. 60208}
\author{D.A. Roberts}
\affil{Department of Physics and Astronomy, Northwestern University,
Evanston, IL. 60208}
\author{Geoff Bower}
\affil{UC Berkeley, Berkeley, CA 94720}
\author{M. Wardle}
\affil{School of Physics, University of Sydney, NSW 2006, Australia}

\begin{abstract}

A new class of OH (1720 MHz) masers unaccompanied by main-line
transitions have recently been discovered (Frail, Goss and Slysh
1994). These masers lie at the interface between supernova remnants (SNRs)
interacting with molecular clouds.  We discuss three new aspects of SN
masers found in the direction toward the Galactic center: (i) the
detection of a new --130 \kms\ OH (1720 MHz) maser in the southern
lobe of the molecular ring at the Galactic center, (ii) the
detection of extended OH (1720 MHz) maser emission from W28
accompanying the compact maser sources and (iii) the detection of
linear polarization of the brightest OH (1720 MHz) maser in SNR
G359.1--0.5.

\end{abstract}

\section{Introduction}

It is increasingly evident that the new class of so-called ``supernova
masers'' is distinguishing itself as an excellent signature of the
interaction of SNR's and molecular clouds. In addition, there is now
considerable theoretical and observational evidence that OH (1720 MHz)
maser spots probe the physical condition of postshock gas which
results from the expansion of a SNR driving a shock into
a molecular cloud as discussed in the conference (e.g., Green 2001).  
The study of a number of SN
masers have led to some of the first measurements of postshock
magnetic field (e.g., Brogan 2001).

There are a total of 20 known supernova masers in the Galaxy, the
largest concentration of which is observed toward the center of the
Galaxy (Yusef-Zadeh \etal 1999). The reason for such a concentration
is not clear but high molecular density ranging between 10$^4$ and
10$^5$ cm$^{-3}$ combined with OH column density of 10$^{16} -
10^{17}$ cm$^{-2}$ (Lockett \etal 1999) as well as a high
concentration of SNRs in this region are needed to produce the larger
number of SN masers observed in the Galactic center region. Here we
discuss some new characteristics of three different SN masers located
between us and the Galactic center.

\section{Sgr A East: the  Blue-shifted OH Feature}

Perhaps the most complicated group of SN masers in the Galaxy is the
one located at the Galactic center. A clumpy molecular ring (also
known as the Circumnuclear Disk or CND), on a scale of 2 to 5 pcs is
circling the Galactic center with a rotational velocity of about 100
\kms\ (see Jackson \etal 1993). On a larger scale of up to 15 pcs, SNR
G0.0+0.0 or Sgr A East is known to be dynamically coupled to the 50
\kms\ molecular cloud.  The detection of several OH (1720 MHz) maser
spots at velocities ranging between 43 and 66 \kms\ surrounding the
Sgr A East shell was first reported in 1996. These observations
supported the picture of the physical interaction of Sgr A East and
the 50 \kms\ molecular cloud (Yusef-Zadeh \etal 1996, 1999). These
studies show another group of OH (1720 MHz) masers to the northeast
lobe of the molecular ring at the velocity of +134 \kms.  The inferred
magnetic field strength along the line of sight due to Zeeman
splitting is between 2 and 5 mG.  The orientations of the magnetic
fields along the line of sight are away from the Earth for all
masers except for the +134 \kms\ CND masers, which are directed toward
the Earth.  The cluster of highly red-shifted OH (1720 MHz) masers
associated with the CND implies that the Sgr A East SNR shell is also
expanding into the molecular ring and is responsible for the
production of the OH (1720 MHz) maser emission behind shock front.
Recent molecular H$_2$ observations of the Sgr A East masers supports
the picture in which OH (1720 MHz) masers are tracing post-shocked gas
(Yusef-Zadeh \etal 2001).  The presence of such a cluster of OH (1720
MHz) masers at +134 \kms\ provides compelling evidence that Sgr A
East is indeed very close to the Galactic center and is interacting
with the molecular ring circling the center of the Galaxy.

An even more striking result was realized when we analyzed the
archival OH (1720 MHz) data taken in 1986 based on A-array VLA
observation of the Galactic center with a spatial and spectral
resolutions of 4.4$''\times3.6''$ and 8.6 \kms,
respectively. This 1986 observation confirmed all the bright OH (1720
MHz) maser spots which were observed about ten years later in 1996.
Understandably the class of SN masers could have been discovered
serendipitously several years earlier than the rediscovery paper by
Frail, Goss and Slysh in (1994) which was published 26 years after the
detection of masers toward SNRs (e.g., Goss 1968).  Another
interesting aspect of the 1986 data is the detection of a highly
blue-shifted OH (1720 MHz) maser line at the velocity of --131.6 \kms\
at $\alpha=17^{\rm h} 42^{\rm m} 28.^{\rm s}03, \delta=-29^\circ 00'
6.''4$.  This maser has not been detected since then in spite of
efforts made recently by M. Goss (private communication). The spectrum
of this new maser has a peak flux density of 75 mJy with the rms noise
of 6.6 mJy beam$^{-1}$. The peak flux is 4 times brighter than the
3$\sigma$ upper limit of 16 mJy from the 1996 epoch
observations. Figure 1a shows the positions of the symmetrically
placed blue- and red-shifted OH masers as crosses and are superposed
on the molecular ring traced by H$_2$ emission (Yusef-Zadeh \etal
2001).  The velocity of the masers is consistent with the sense of
rotation of the molecular ring.  We believe the $\pm$132 \kms\
velocity features are signifying the true rotational velocity of the
molecular gas orbiting the Galactic center. This is because the
largest amplification occurs in the direction of largest velocity
coherence which tends to be perpendicular to the shock front as it
accelerates the gas in the transverse direction.  This implies that
the velocity of masers reflects the systemic motion of the molecular
cloud.  If so, the mass enclosed within a radius of about 45$''$ (1.8
pc at the distance of 8.5 kpc) from Sgr A$^*$ is estimated to be about
7$\times10^6$ \msol using Keplerian motion.  We can also interpret
that median velocity of the OH (1720 MHz) masers in the ring is
consistent with no radial motion of the Local Standard of Rest with
respect to the Galactic center.

\section{W28: Extended Maser Emission}

Studies of the inner several degrees of the Galactic plane toward the
Galactic center detected \emph{extended} OH (1720 MHz) emission from
G359.1--0.5, G357.7+0.3 (The Square Nebula) and G357.7--0.1 (the
Tornado Nebula) (Yusef-Zadeh et al., 1995; 1999).  The largest
extended OH (1720 MHz) emission arises from G357.7+0.3 distributed over
an angular size of $> 15'$ corresponding to 27 pc at the distance of
6.4 kpc (Leahy 1989).  More recently, we observed W28 at the distance
of 2 kpc and detected new extended OH (1720 MHz) features coincident
with the compact maser spots at similar radial velocities. Figure 1b
shows contours of extended emission superposed on the $\lambda$20cm
continuum image of W28 with a resolution of 28$''\times16''$. It is
clear that the extended OH (1720 MHz) emission follows the continuum
morphology.

These features are probably low-gain masers that co-exist with the
bright, compact masers, as has been argued in the case of G359.1--0.5
(Yusef-Zadeh, Uchida \& Roberts 1995).  Extended OH (1720 MHz)
emission may in principle be produced by thermal foreground objects in
a manner analogous to the widespread Galactic OH (1720 MHz) emission
observed in the Galactic plane (Turner 1982), but the the narrow line
widths (a few \kms) and the similarity in the 
kinematics and the distribution of extended features and compact 
maser spots all suggest
that the extended OH (1720 MHz) emission is nonthermal.
The significance  of extended emission from SN masers 
is that it is probably  tracing the region of shocked molecular gas due
to shock chemistry enhancing the abundance 
of OH molecule (Wardle 1999).
 Recent work by Wardle et al. (2001) indicates the
morphology of 
extended OH (1665 and 1667 MHz) line 
in absorption is similar to that of extended OH (1720 MHz) line 
in emission but with broad linewidths. The distribution of OH (1720 MHz) line in emission
and OH(1665 MHz) in absorption 
associated W28 also is very similar 
to that of shocked CO cloud identified by Arikawa \etal (1999). 

\section{G359.1--0.5: The Linear Polarization}

G359.1--0.5 is identified as a shell-type SNR with a ring of of CO
cloud surrounding it OH (1720
MHz) observations of this source indicated several compact maser
sources at the edge of the remnant having velocities $\sim$--5 \kms\
(Yusef-Zadeh \etal 1995). We have recently observed this source with
the VLA in its A configuration to search for polarized emission from
OH (1720 MHz) masers. Figure 2a shows the spectra of all four Stokes parameters toward
the brightest maser (A). Circular polarization due to Zeeman splitting
is fitted and the inferred magnetic field along the line of sight is
estimated to be +561 $\pm$ 32 $\mu$G. The linearly polarized
emission at this wavelength is also detected with a fractional
polarization of about 8\%. The distribution of the electric field as
noted in Fig. 2b is is seen predominantly in
the direction perpendicular to the edge of the remnant at a position angle
of abut --70$^\circ$. Theoretical modeling of  the magnetic field 
direction suggests that the B field 
either aligns with the polarization direction or runs 
perpendicular to it (Elitzur 1996).
It is 
interesting that recent observations of the S(1) 1--0 H$_2$ line
emission from the edge of the remnant toward the A maser indicate an
elongated H$_2$ feature with the size of 90$''\times15''$
(length$\times$width) directed tangential to the shell with  a
PA $\approx$40$^\circ$ (Lazendic \etal
2001). The H$_2$ emission is thought to arise from a shock front which
is driven by the expansion of the remnant. Assuming that the magnetic
field lines are oriented along the shock front and that the OH (1720
MHz) maser and H$_2$ emission arise from the cooling gas behind a
C-type shock, the direction of the magnetic field is then aligned with
the long axis of the H$_2$ feature. This implies that the B
field ambiguity is resolved and and that 
it  runs perpendicular to  the polarization vectors of linearly
polarized emission, as noted in Fig. 2b.


Acknowledgments: We thank M. Elitzur and 
W. Watson for useful discussions.


\begin{figure}

\caption{Left panel (Fig. 1a) shows the blue- and red-shifted OH masers as crosses superposed on the
Galactic center molecular ring traced by H$_2$ emission (Yusef-Zadeh \etal 2001).
Right panel (Fig. 1b) shows contours of extended emission superposed on
the $\lambda$20cm continuum image of W28 with a resolution of
28$''\times16''$.}

\end{figure}

\begin{figure}

\caption{Left panel (Fig. 2a) shows the spectra of all four Stokes parameters toward
the brightest maser (A).  The distribution of the electric field is
shown in the right panel (Fig. 2b) and is seen predominantly in the
direction perpendicular to the edge of the remnant.}

\end{figure}

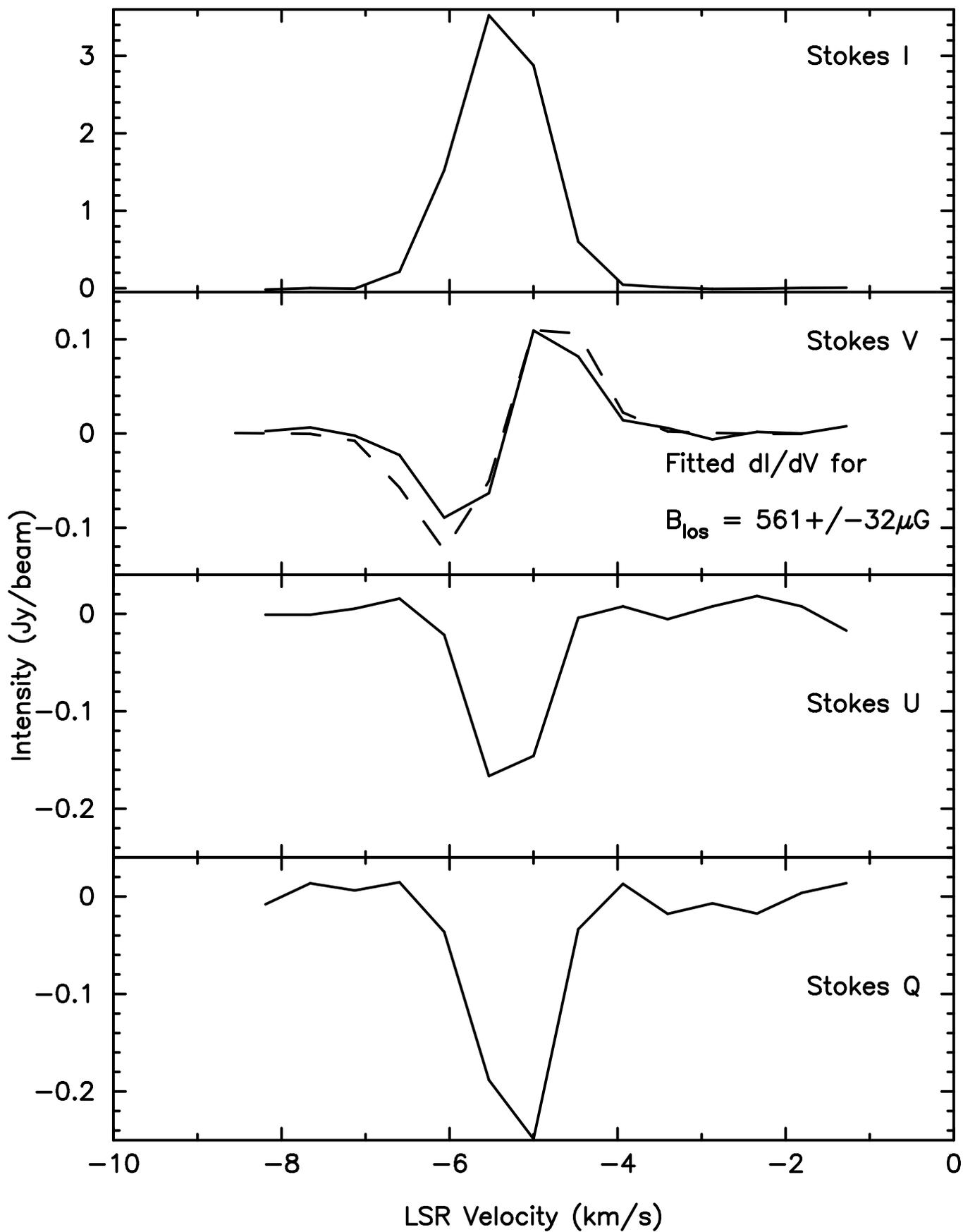
\end{document}